\documentclass[final,5p,times,sort,compress]{elsarticle}
\usepackage{lmodern}
\usepackage[normalem]{ulem}
\usepackage{amssymb}
\usepackage{amsmath}
\usepackage{lipsum}
\usepackage[colorlinks=true, linkcolor=blue, citecolor=blue, urlcolor=blue]{hyperref}
\usepackage[nameinlink]{cleveref}
\crefname{figure}{Figure}{Figures}
\Crefname{figure}{Figure}{Figures}
\crefname{equation}{Eq.}{Eqs.}
\Crefname{equation}{Eq.}{Eqs.}
\crefname{table}{Table}{Tables}
\Crefname{table}{Table}{Tables}

 \usepackage{tikz}
 \usetikzlibrary{positioning, fit, shapes.geometric}
 \usepackage{caption}
\usepackage{booktabs}

\usepackage{verbatim}

\usepackage{microtype}

\usepackage[htt]{hyphenat}

\usepackage{soul}
\usepackage{xcolor}

\newcommand{\added}[1]{#1}

\usepackage{listings}
\usepackage{xcolor}

\usepackage{tabularx}
\usepackage{multirow}

\usepackage{cuted}
\usepackage{capt-of}
\journal{SoftwareX}

\begin{document}
\renewcommand{\labelenumii}{\arabic{enumi}.\arabic{enumii}}

\begin{frontmatter}
 


\title{GEMSS: A C++ Library for Multi-Sphere Modeling in DEM Simulations}
\author{A. Moradian}
\author{F. Buchele}
\author{P. Müller}
\author{T. Pöschel}
\address{Lehrstuhl f\"ur Multiskalensimulation, Friedrich-Alexander-Universit\"at Erlangen-N\"urnberg, Germany}

\begin{abstract}
\textit{GEMSS (GEnerator of Multi-Sphere Shapes) converts 3D surface meshes or voxel grids into multi-sphere representations of granular particles using 
the recently published MSS algorithm.
It computes key physical properties required for discrete element method (DEM) and general multibody dynamics simulations, including particle volume, center of mass, and principal moments and axes of inertia. Implemented as a header-only C++ library, GEMSS is easily integrated into DEM and molecular dynamics frameworks. The library has been integrated into MercuryDPM, which enables on-the-fly generation of multi-sphere particles directly within the simulation loop.}
\end{abstract}

\end{frontmatter}
\vspace{-0.8\baselineskip}
\begin{strip}
\vspace{-1.0\baselineskip}
%
\end{strip}

\section{Introduction}

Accurate representation of individual particle shapes is often essential for describing packing structure, contact interactions, and flow behavior in granular systems, using the Discrete Element Method (DEM) or Molecular Dynamics (MD) \cite{Zhao:2017, Cleary:2008, Lu:2015, Salerno:2018, zhao.2023, Ulusoy:2023, Jiang2025, Luo2024, Wang2024, Nie2020}.

Among the available methods for representing particle shape, the multi-sphere model \cite{Poeschel.1993,Buchholtz.1994,Buchholtz.1996} provides a good compromise between computational cost and realistic particle shape description \cite{Kruggel:2008, Markauskas:2010}. It is widely used in DEM simulations of granular systems involving non-spherical particles \cite{Favier:1999, Thornton:2000, Bradshaw:2004,  Amberger:2012, Garcia.2009, Markauskas:2010,  Lu:2015, Zhou:2016, zhang:2021, Yuan:2023, Chen:2025}.

\begin{figure}[htbp]
    \centering
    \begin{minipage}{0.48\linewidth}
        \centering
        \includegraphics[width=\linewidth]{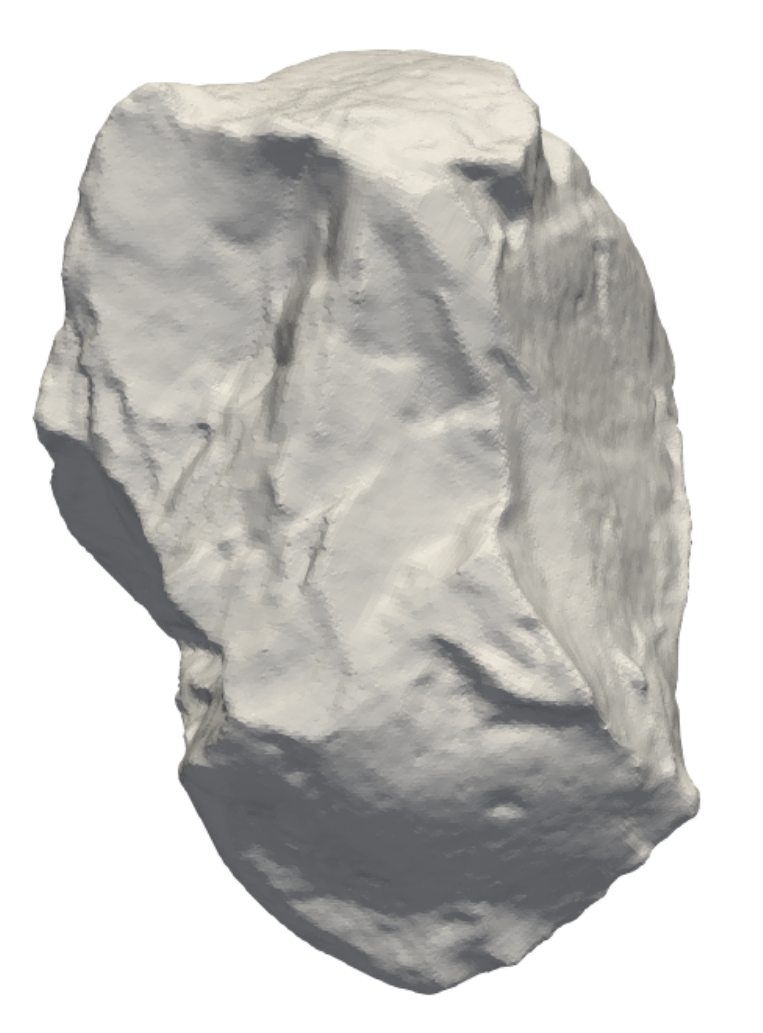}
    \end{minipage}\hfill
    \begin{minipage}{0.48\linewidth}
        \centering
        \includegraphics[width=\linewidth]{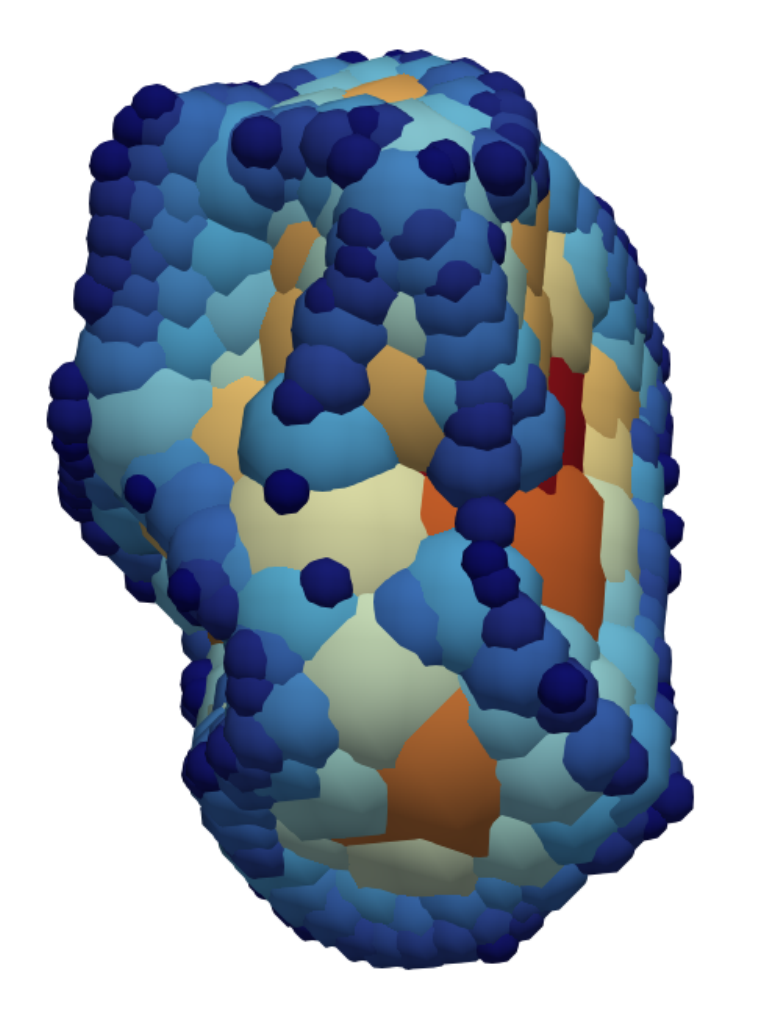} 
    \end{minipage}
    \caption{A complex-shaped particle described by a surface mesh and its multi-sphere model for $n=360$. Colors indicate the radii of the individual spheres. The particle corresponds to a \textit{Hamburg sand} grain taken from the \textit{Sand Atlas} \cite{Milatz.2021} }
    \label{fig:firstExample}
\end{figure}

To represent a granular particle of a given shape $\mathcal{S}$ using the multi-sphere model, a set of $n$ spheres is arranged such that their union 
\begin{equation}
   \tilde{\mathcal{S}} \equiv \bigcup_{i=0}^{n-1} \tilde{\mathcal{S}}_i(R_i, \vec{r}_i)\,, 
   \label{eq:sphereRep}
\end{equation}
approximates the target shape $\mathcal{S}$ as closely as possible. For a fixed number of spheres, $n$, a generation algorithm determines the radii $R_i$ and positions $\vec{r}_i$ ($i=0,\dots,n-1$) to minimize the deviation between $\mathcal{S}$ and $\tilde{\mathcal{S}}$. \Cref{fig:firstExample} shows an example of a complex shaped particle and its multi-sphere representation. 

Several algorithms have been proposed for generating multi-sphere particle models from particle shape descriptions  \cite{Wang:2007, Lu:2007, Ferellec:2008, Ferellec:2010, Taghavi:2011, Gao:2012, Li:2015, Zheng:2016, Zheng:2017, Haeri:2017, Westbrink:2017, Tian:2018, Katagiri:2018, Mede:2018, Yuan:2019, Zhou:2019, Suhr:2020, Nie:2022, Li:2022, Xiong:2023, Ma:2023}, a critical comparison can be found in Ref. \cite{Fathipour:2025}. Some of these algorithms have been implemented and made available to the scientific community \cite{Favier:1999, Angelidakis.2021, canbolat.2025, Multisphere.2026}, however, none of them are provided as library components that can be executed on-the-fly, during a running DEM simulation. Instead, they are implemented as standalone programs in interpreted languages such as Python or \textsc{Matlab}. Consequently, updates to the multi-sphere representation of particles within a simulation require a fragmented workflow: the user must generate the multi-sphere representation of a particle, compute the required physical parameters, and pass the  data to the DEM framework. This workflow limits the simulation of phenomena that require changes in particle shape, such as comminution, abrasive wear, and plastic deformation.

The GEnerator of Multi-Sphere Shapes (GEMSS) addresses these limitations. Building upon the Multi-Sphere Shape (MSS) algorithm \cite{Buchele.2026}, GEMSS combines multi-sphere generation and physical property calculations into a unified workflow. By shifting to runtime generation, it reduces the computational overhead of particle model generation. This allows users to generate multi-sphere particles directly within the active simulation loop.

As a header-only C++17 library, GEMSS can integrate directly into established frameworks such as MercuryDPM \cite{Weinhart.2020}, YADE \cite{Angelidakis.2024} and LAMMPS \cite{Plimpton.1995}. It requires Eigen, libigl, and edt, with optional support for shared memory parallelization using OpenMP.

Integration involves placing the header files into a host project. A CMake configuration is provided that automates dependency management. The complete source code, \added{along with the test scripts and input data required to reproduce the examples presented in this manuscript}, is freely available on GitHub, see \cite{Moradian.2026}.


\section{Multi-sphere modeling with GEMSS}
\subsection{General outline}

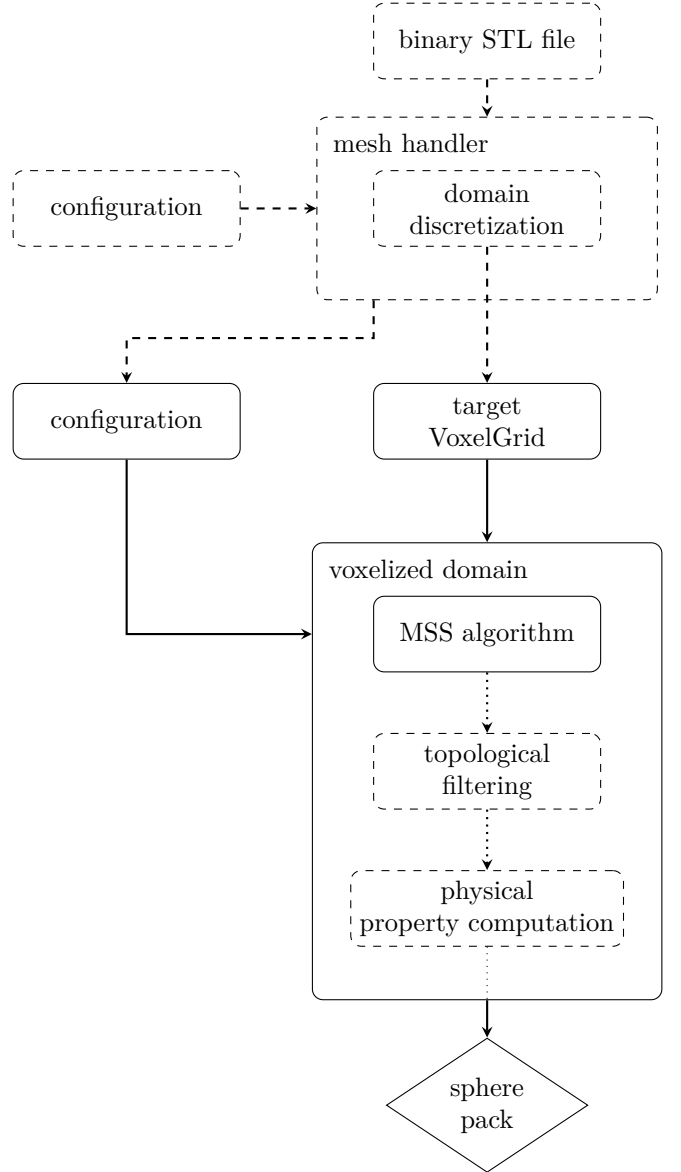
\begin{figure}[t]
    \centering
    \begin{tikzpicture}[
        process/.style={rectangle, minimum width=3cm, minimum height=1cm, text centered, draw=black, rounded corners, align=center},
        decision/.style={diamond, minimum width=2.5cm, minimum height=1cm, text centered, draw=black, align=center, aspect=1.5},
        arrow/.style={thick,->,>=stealth}
    ]

    \node (stl) [process, dashed] {binary STL file};
    
    \node (disc) [process, dashed, below=1.2cm of stl] {domain\\discretization};
    
    \node (grid) [process, below=1.8cm of disc] {target\\VoxelGrid};
    
    \node (mss) [process, below=1.8cm of grid] {MSS algorithm};
    \node (topo) [process, dashed, below=0.8cm of mss] {topological\\filtering};
    \node (phys) [process, dashed, below=0.8cm of topo] {physical\\property computation};
    
    \node (sphere) [decision, below=1.2cm of phys] {sphere\\pack};

    \node (mesh_box) [draw=black, dashed, rounded corners, fit=(disc), inner sep=0.5cm, inner ysep=0.7cm, minimum width=4.5cm] {};
    \node [anchor=north west, xshift=0.1cm, yshift=-0.1cm] at (mesh_box.north west) {mesh handler};

    \node (vox_box) [draw=black, rounded corners, fit=(mss) (topo) (phys), inner sep=0.5cm, inner ysep=0.7cm, minimum width=4.5cm] {};
    \node [anchor=north west, xshift=0.1cm, yshift=-0.1cm] at (vox_box.north west) {voxelized domain};

    \node (config1) [process, dashed, left=1cm of mesh_box.west |- disc] {configuration};
    
    \node (config2) [process, at={(config1 |- grid)}] {configuration};

    
    \draw [arrow, dashed] (stl) -- (mesh_box.north -| stl);
    \draw [arrow, dashed] (config1) -- (mesh_box.west |- config1);

    \draw [arrow, dashed] (disc) -- (grid);
\draw [arrow, dashed] ([xshift=-1.5cm]mesh_box.south) -- +(0,-0.5) -| (config2.north);
    \draw [arrow] (grid) -- (vox_box.north -| grid);
    \draw [arrow] (config2.south) |- (vox_box.west |- mss); 

    \draw [arrow,dotted] (mss) -- (topo);
    \draw [arrow,dotted] (topo) -- (phys);
    \draw [dotted] (phys) -- (vox_box.south -| sphere);
    \draw [arrow] (vox_box.south -| sphere) -- (sphere);
    \end{tikzpicture}
    \caption{Architecture of GEMSS. Dashed boxes indicate optional steps}
    \label{fig:architecture_vertical}
\end{figure}

The architecture of \texttt{GEMSS} is illustrated in \cref{fig:architecture_vertical}. Starting from the particle description by a surface mesh (STL format), the internal data flows through three stages: shape discretization, multi-sphere model generation, and physical property computation. The first stage can be skipped if the particle description is already given by a voxel grid.

\subsubsection*{Core Data Abstractions}
The library relies on three main structures within the \texttt{GEMSS} namespace:
\begin{itemize}
\item \textbf{\texttt{MultisphereConfig}:} A unified configuration structure for solver settings, including grid resolution, termination criteria, and optional features.
    \item \textbf{\texttt{VoxelGrid<T>}:} A voxel representation of the shape, defined as a binary mask or a distance field. 
    \item \textbf{\texttt{SpherePack}:} The primary output consists of the multi-sphere representation (sphere centers and radii) and, optionally, the particle’s physical properties.
\end{itemize}

\subsubsection*{Public Interface}
The user interface is provided through a single header. \cref{fig:code_snippet} shows the complete code required to load a shape, configure the solver, and generate a multi-sphere particle.
\begin{figure}[t]
    \centering
    \begin{lstlisting}
#include "GEMSS-interface.h"
#include <iostream>

using namespace GEMSS;

int main() {
  // 1. load the target shape
  STLMesh mesh = 
     load_mesh("complex_particle.stl");

  // 2. define the 
  //approximation configuration
  MultisphereConfig config;
  // voxel grid resolution
  config.div = 150;      
  // 99% volume overlap target
  config.precision_target = 0.99;   
  // hard limit on the number of spheres
  config.max_spheres = 100;
  // compute CoM & Inertia 
  // 1 = from approximation
  // 2 = from target grid
  config.compute_physics = 1;         

  // 3. execute the approximation
  SpherePack sp =
        multisphere_from_mesh(mesh, config);

  // 4. access the physical metadata
  std::cout<< "Volume: "<< sp.volume<< "\n"
         << "Center of Mass:\n" 
         << sp.center_of_mass << "\n"
         << "Inertia:\n" 
         << sp.inertia_tensor << std::endl;

  return 0;
}
    \end{lstlisting}
    \caption{A minimal C++ example demonstrating the generation of a multi-sphere model using the \texttt{GEMSS} API.}
    \label{fig:code_snippet}
\end{figure}

GEMSS accepts binary STL files or VoxelGrids as input. Approximation constraints are defined via the \texttt{MultisphereConfig} (lines 13--23). A single function call (line 26) executes the discretization, MSS algorithm, and postprocessing computations within a voxelized domain. The resulting \texttt{SpherePack} object (lines 30--34) stores all physical properties, enabling direct mapping to a host DEM solver or export to VTK and CSV formats for visualization.

\subsection{GEMSS functionalities}

\subsubsection*{Adaptive Shape Fidelity}

\begin{figure}[htbp]
    \centering
    \begin{tikzpicture}[>=stealth, thick]
        \def\R{1.5}     
        \def\d{2.5}     
        \def\halfD{1.25} 
        
        \pgfmathsetmacro{\yInt}{sqrt(\R*\R - \halfD*\halfD)}
        \pgfmathsetmacro{\gap}{\R - \yInt}

        \draw[fill=gray!10] (-\halfD, 0) circle (\R);
        \draw[fill=gray!10] (\halfD, 0) circle (\R);

        \filldraw (-\halfD, 0) circle (2pt) node[below] {$C_1$};
        \filldraw (\halfD, 0) circle (2pt) node[below] {$C_2$};

        \draw[<->, dashed] (-\halfD, 0) -- (\halfD, 0) node[midway, below] {$d$};

        \draw[->] (\halfD, 0) -- ++(45:\R) node[midway, below right] {$R$};

        \draw[dashed, blue] (-\halfD - 1, \R) -- (\halfD + 1, \R) node[right, text=black] {tangent plane};

        \draw[<->, red, ultra thick] (0, \yInt) -- (0, \R) node[midway, right] {\!\!$\epsilon$};
        
        \filldraw[red] (0, \yInt) circle (1.5pt);

    \end{tikzpicture}
    \caption{Schematic of roughness description. The scallop height, $\epsilon$, depends on the sphere radius $R$ and the center-to-center distance $d$.}
    \label{fig:scallop_height}
\end{figure}
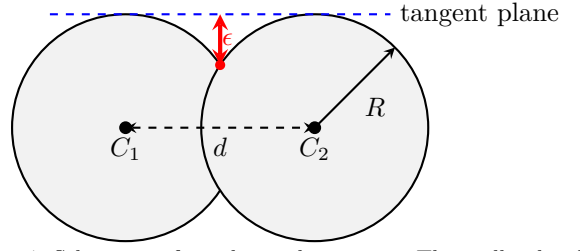

In multi-sphere modeling, discrete approximations introduce a geometric artifact commonly referred to as \textit{artificial roughness} or \textit{bumpiness} \cite{Ferellec.2010,Angelidakis.2021,Soltanbeigi.2017,Fathipour.2018}. 
\texttt{GEMSS} quantifies and controls this artificial roughness by applying the geometric concept of \textit{scallop height} $\epsilon$ \cite{Suresh1994, Choi1998}, defined as the maximum gap depth between intersecting spheres and their common external tangent plane (\cref{fig:scallop_height}). 
To promote a uniform $\epsilon$ value across the approximated shape, the center-to-center distance $d$ between intersecting spheres scales proportionally with the square root of the local radius:
\begin{equation} \label{relativedistance}
    d_{\min} = k\sqrt{R}\,,
\end{equation}
where $k$ is a distance scaling factor selected by the user. 

\subsubsection*{Continuity check}
Owing to its greedy nature, the MSS algorithm \cite{Buchele.2026} may yield disconnected particle representations under strict limits on the number of spheres, as it prioritizes large-scale features over thin connecting regions. To mitigate this effect, \texttt{GEMSS} provides an optional \textit{topological filtering} step that removes small, isolated sphere clusters. 

\added {The occurrence of isolated clusters depends on particle morphology and the imposed sphere limit. For example, compact particles like Hamburg sand rarely require filtering for $n > 10$, whereas highly irregular lunar particles often produce disconnected clusters. The application of this filter is optional, as disjointed representations (e.g., the two disconnected ends of a dumbbell) may provide the most accurate volumetric and inertial model under strict sphere limits. Disconnected clusters can be simulated as rigid clumps in DEM; however, they permit unphysical particle interpenetration in the empty regions between clusters. Conversely, filtering removes geometric extremities, reducing the overall volume and similarly permitting interpenetration in regions where the physical particle's volume was truncated.}

\subsubsection*{Physical property computation}

Calculating a particle's mass, center of mass, and inertia tensor is necessary for solving rigid-body dynamics \cite{Ferellec.2010}. \texttt{GEMSS} computes these properties via discrete summation over a voxel grid  \cite{Ostanin.2024}.  Users can choose to perform this calculation using either the voxelized multi-sphere model or the original target shape.




\section{Illustrative examples}


\subsection{Validation of physical properties}

We apply \texttt{GEMSS} to a set of representative target bodies and evaluate how accurately their properties are captured by the corresponding multi-sphere approximations, see \cref{tab:comptable}. The results show relative errors under 2\% for volume and under 3\% for principal moments of inertia. For the center of mass, the Euclidean distance between the predicted and actual positions was normalized by the cubic root of the particle volume, used here as a characteristic object size. The resulting maximum error is 0.59\%. 

We compare the principal moments of inertia of the target object and its multi-sphere model to ensure the results are independent of coordinate orientation. The inertia tensor is computed from the voxelized representation of the multi-sphere model and diagonalized by an eigendecomposition. We then sort the resulting principal moments of inertia by magnitude and compare them with the exact principal moments of inertia of the target object, sorted in the same way. The percentage reported in \cref{tab:comptable} corresponds to the maximum relative deviation between a principal moment of inertia of the multi-sphere model and its exact counterpart.

\begin{table*}[!htbp]
    \centering
    \small 
\caption{
Comparison of the physical properties of multi-sphere models with the exact values for some elementary bodies (sphere, cube, cylinder, cone, hemisphere).
For volume and principal moments of inertia, the deviation is quantified by the relative difference between the exact values and the corresponding values of the multi-sphere model. For the center of mass, the deviation is quantified by the Euclidean distance between the centers of mass of the target shape and the multi-sphere model, normalized by the cubic root of the volume. In all cases, the volume errors are below 2\%, the moment of inertia errors are below 3\%, and the center-of-mass deviations are below 0.59\%. 
}
\label{tab:comptable}
    \renewcommand{\arraystretch}{1.4} 
    
    \begin{tabularx}{\textwidth}{@{} >{\raggedright\arraybackslash}p{2.8cm} l >{\centering\arraybackslash}X >{\centering\arraybackslash}X c @{}}
        \toprule
        \textbf{} & \textbf{parameter} & \textbf{exact value} & \textbf{multi-sphere model} & \textbf{deviation} \\ 
        \midrule
       \multirow{3}{*}{\includegraphics[width=2.5cm,bb=104 100 263 260,clip]{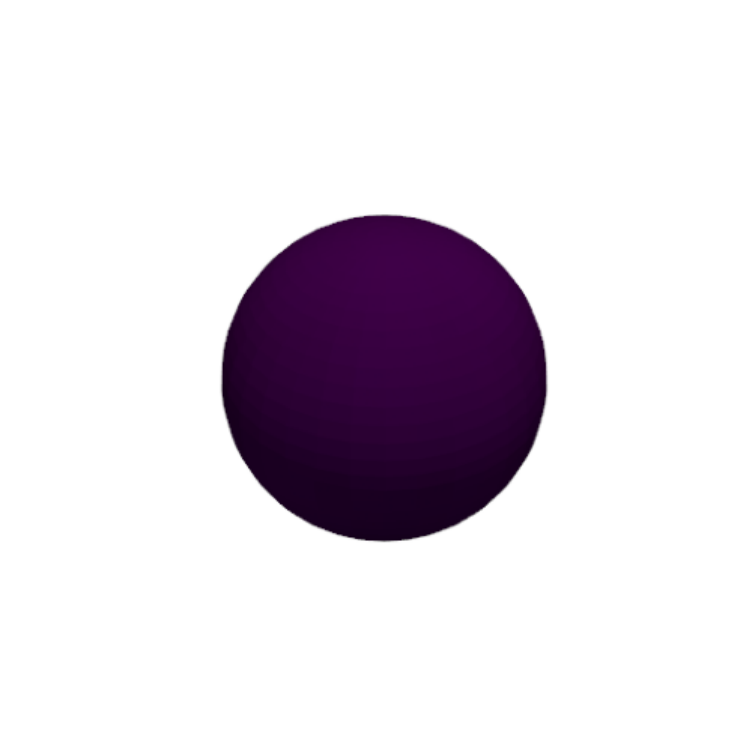}}
       & number of spheres & & 1 \\
       & Dice coefficient & & 0.999\\
       & volume  & 268.08 & 267.95 & 0.05\% \\
        & center of mass & $[5, 5, 5]$ & $[5, 5, 5]$ & 0\% \\
        & moments of inertia & $[1715.73, 1715.73, 1715.73]$ & $[1714.48, 1714.48, 1714.48]$ & 0.07\% \\
        \midrule
       \multirow{3}{*}{\includegraphics[width=2.5cm,bb=74 58 293 288,clip]{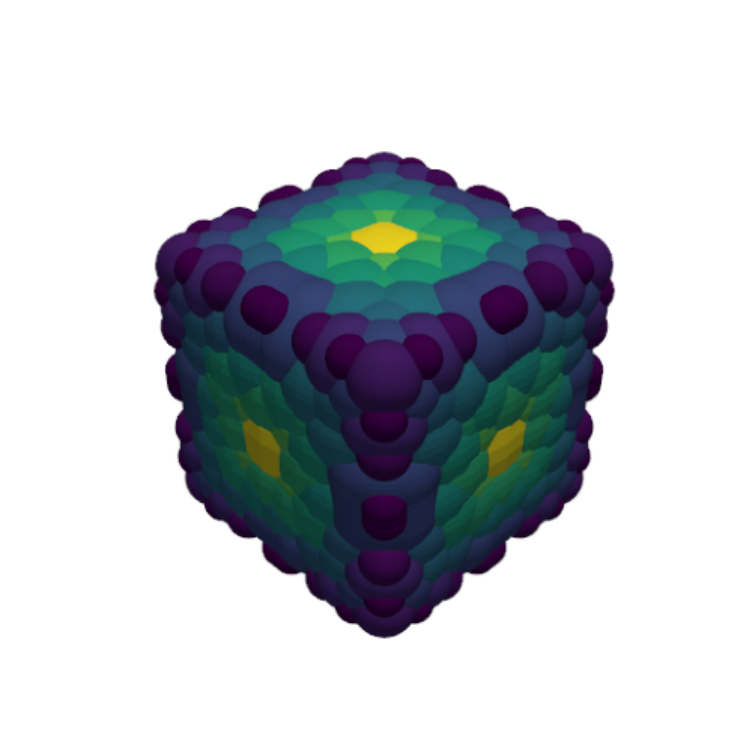}}
       & number of spheres & & 310 \\
       & Dice coefficient & & 0.958\\
        & volume  & 64 & 64.41 & 0.64\% \\
        & center of mass     & $[4.97, 4.97, 4.97]$ & $[4.97, 4.97, 4.97]$ & 0\% \\
        & moments of inertia & $[170.67, 170.67, 170.67]$ & $[170.06, 170.06, 170.06]$ & 0.36\% \\
        \midrule
%
        \multirow{3}{*}{\includegraphics[width=2.4cm,bb=74 62 293 310,clip]{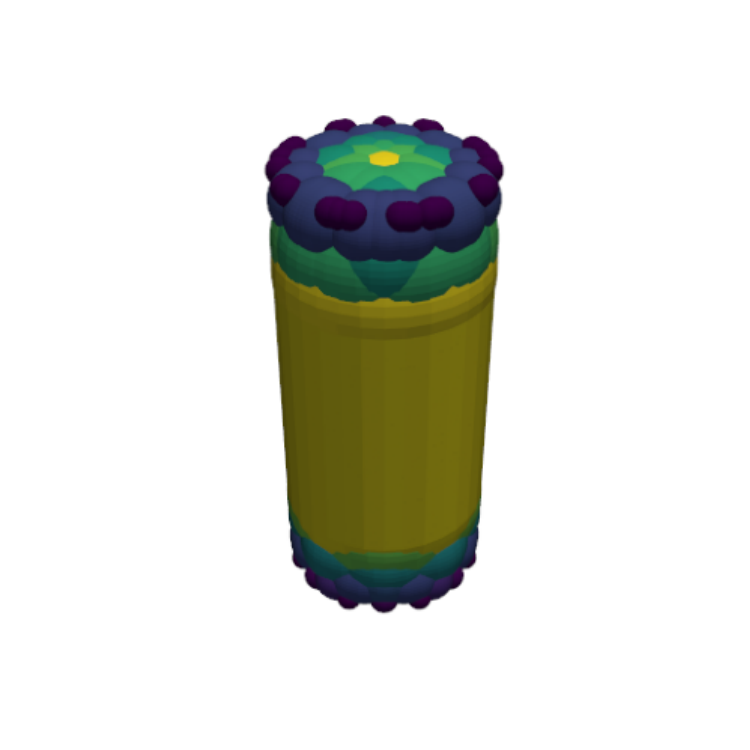}}
       & number of spheres & & 273 \\
       & Dice coefficient & & 0.986\\
       & volume  & 113.1 & 112.39 & 0.63\% \\
        & center of mass     & $[5, 5, 4.97]$ & $[5, 5, 4.97]$ & 0\% \\
        & moments of inertia & $[876.5, 876.5, 226.19]$ & $[  862.21,861.94, 223.44]$ & 1.66\% \\
        \midrule
%
        \multirow{3}{*}{\includegraphics[width=2.8cm,bb=100 115 270 270,clip]{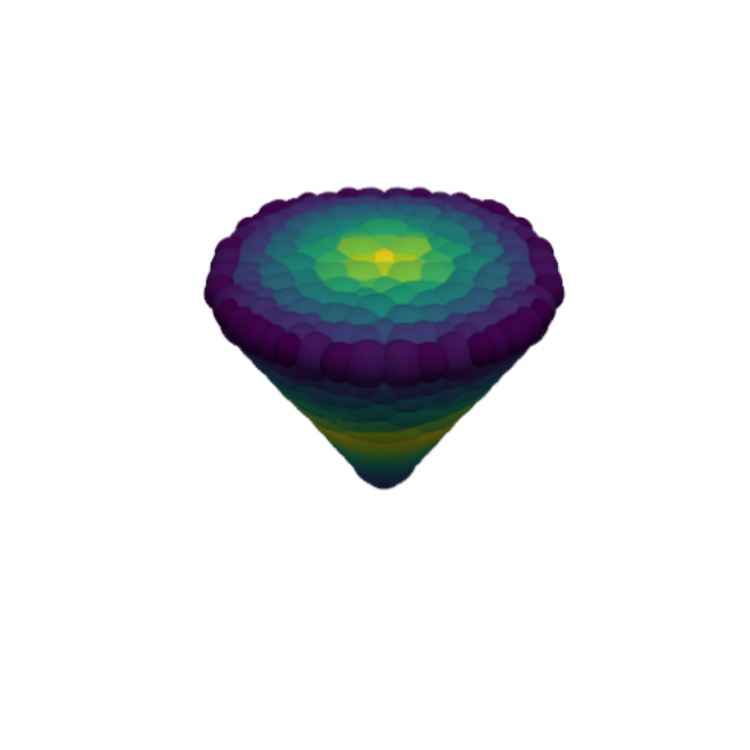}}
        & number of spheres & & 223 \\
       & Dice coefficient & & 0.958\\
        & volume  & 46.65 & 46.68 & 0.06\% \\
        & center of mass & $[5, 5, 6.21]$ & $[5, 5, 6.23]$ & 0.59\% \\
        & moments of inertia & $[105.85, 105.85, 125.96]$ & $[104.06, 104.12, 121.72]$ & 3.37\% \\
        \midrule
 %
         \multirow{3}{*}{\includegraphics[width=2.8cm,bb=95 80 270 235,clip]{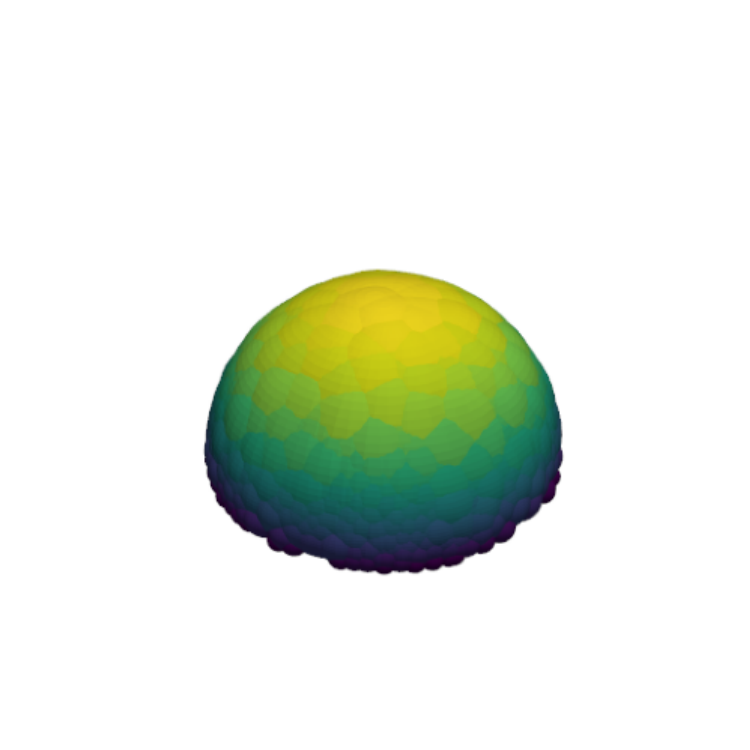}}
         & number of spheres & & 626 \\
       & Dice coefficient & & 0.975\\
        & volume  & 134.04 & 136.68 & 1.97\% \\
        & center of mass     & $[5, 5, 6.5]$ & $[5, 5, 6.5]$ & 0\% \\
        & moments of inertia & $[556.27, 556.27, 857.86]$ & $[567, 569.98, 869.62]$ & 2.47\% \\
        \bottomrule
        
    \end{tabularx}
\end{table*}


\subsection{Accuracy of the model description}

We assess the accuracy of the multi-sphere model using the Dice similarity coefficient \cite{Dice.1945}, which quantifies the mutual intersection volume between the target shape $\mathcal{S}$ and the multi-sphere approximation $\tilde{\mathcal{S}}$:
\begin{equation}
    D \equiv \frac{2 |\mathcal{S} \cap \tilde{\mathcal{S}}|}{|\mathcal{S}| + |\tilde{\mathcal{S}}|}
    \label{eq:Dice}
\end{equation}
where $|\cdot|$ denotes the volume enclosed by a given shape. The accuracy depends on the user-defined distance scaling factor (\cref{relativedistance}) and the number of spheres. When not constrained by the number of spheres, the scaling factor should be less than the square root of the minimum allowed sphere radius. This minimum radius is typically determined by the computational constraints of the DEM simulation.

\begin{figure}[htbp]
        \centering
    \includegraphics[width=1.0\linewidth]{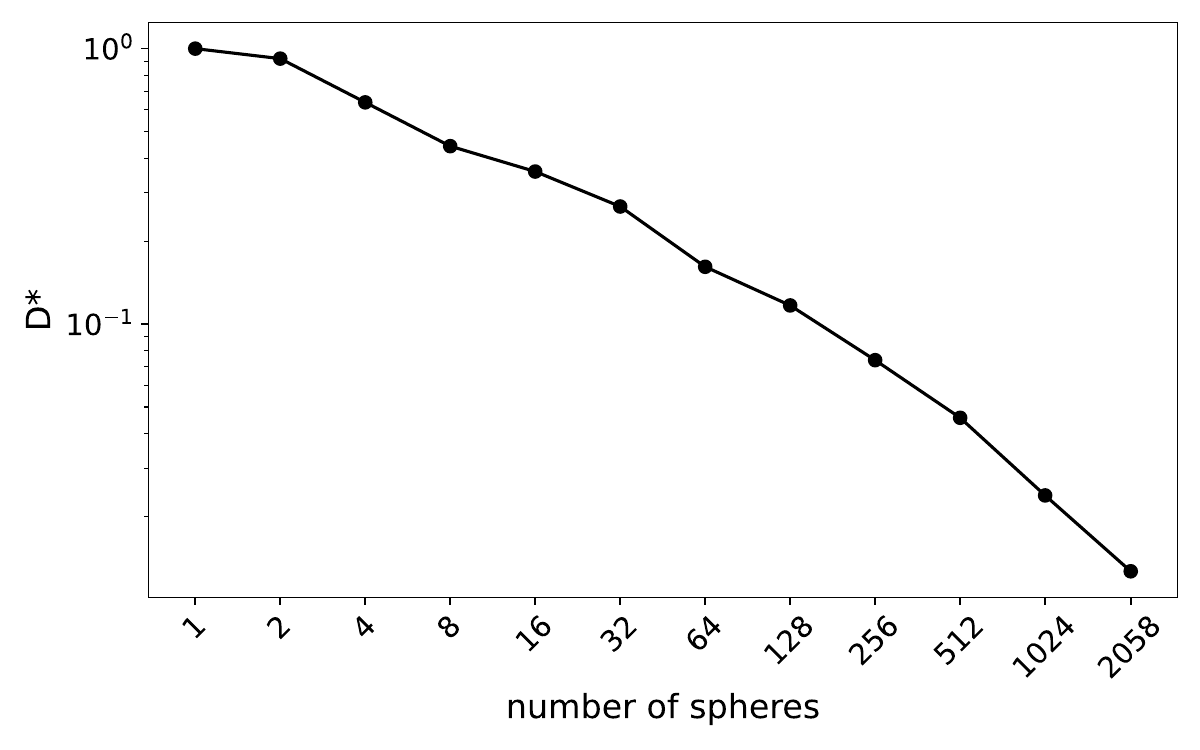}
\caption{Scaled Dice coefficient, $D^\ast$ (\cref{eq:ScaledDice}), as a function of the number of spheres for multi-sphere models of the particle shown in \cref{fig:firstExample}, for k=2.}
\label{fig:boxPrecision}
\end{figure}

\cref{fig:boxPrecision} shows the linearly scaled Dice coefficient,
\begin{equation}
    D^\ast(i) = \frac{1 - D(i)}{1 - D(1)}\,,
    \label{eq:ScaledDice}
\end{equation}
as a function of the number of spheres for the example of the Hamburg sand particle from \cref{fig:firstExample}.  

\begin{figure}[htbp]
    \centering
    \begin{minipage}{0.48\linewidth}
        \centering
        \includegraphics[width=\linewidth]{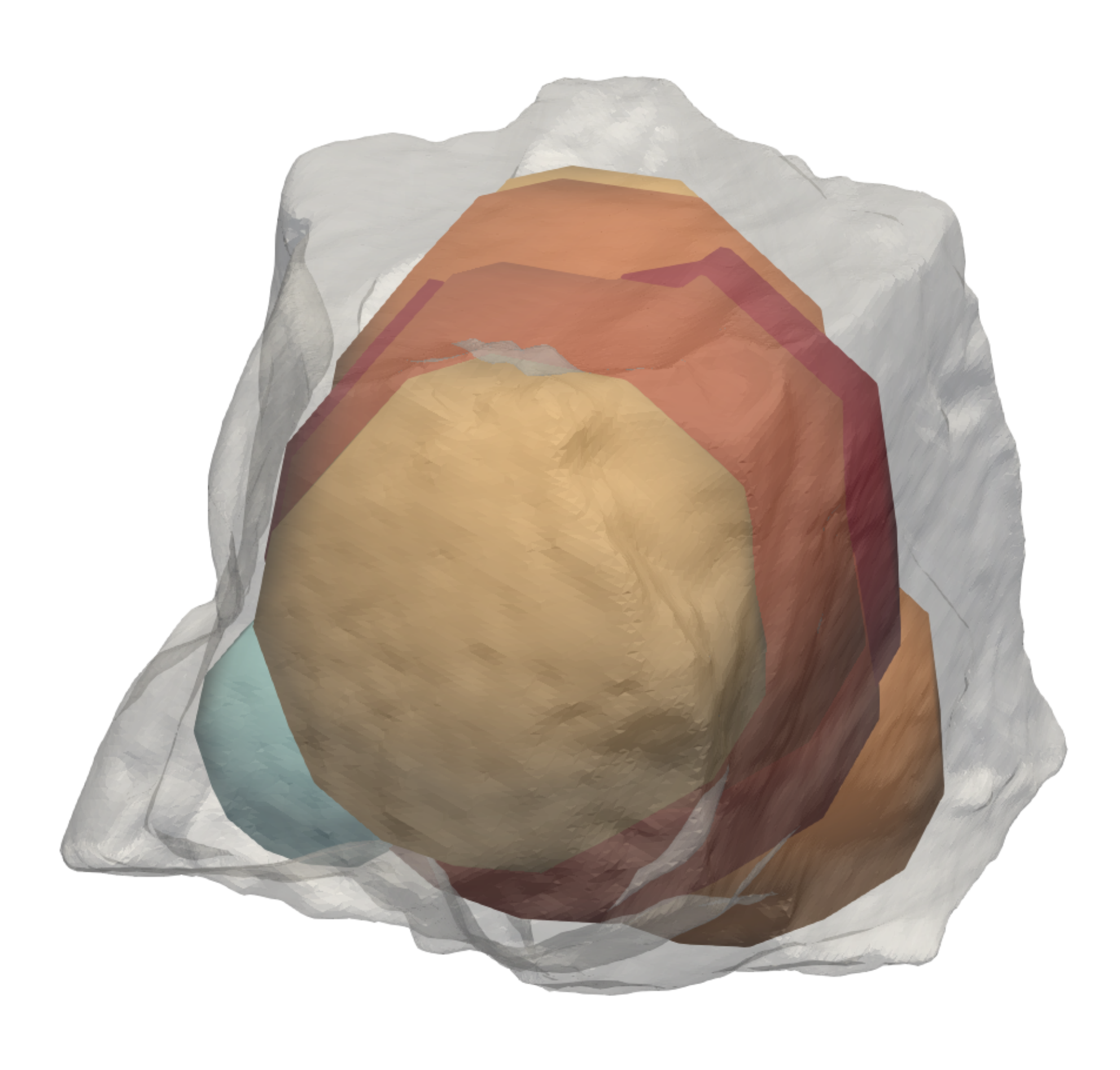}
    \end{minipage}\hfill
    \begin{minipage}{0.48\linewidth}
        \centering
        \includegraphics[width=\linewidth]{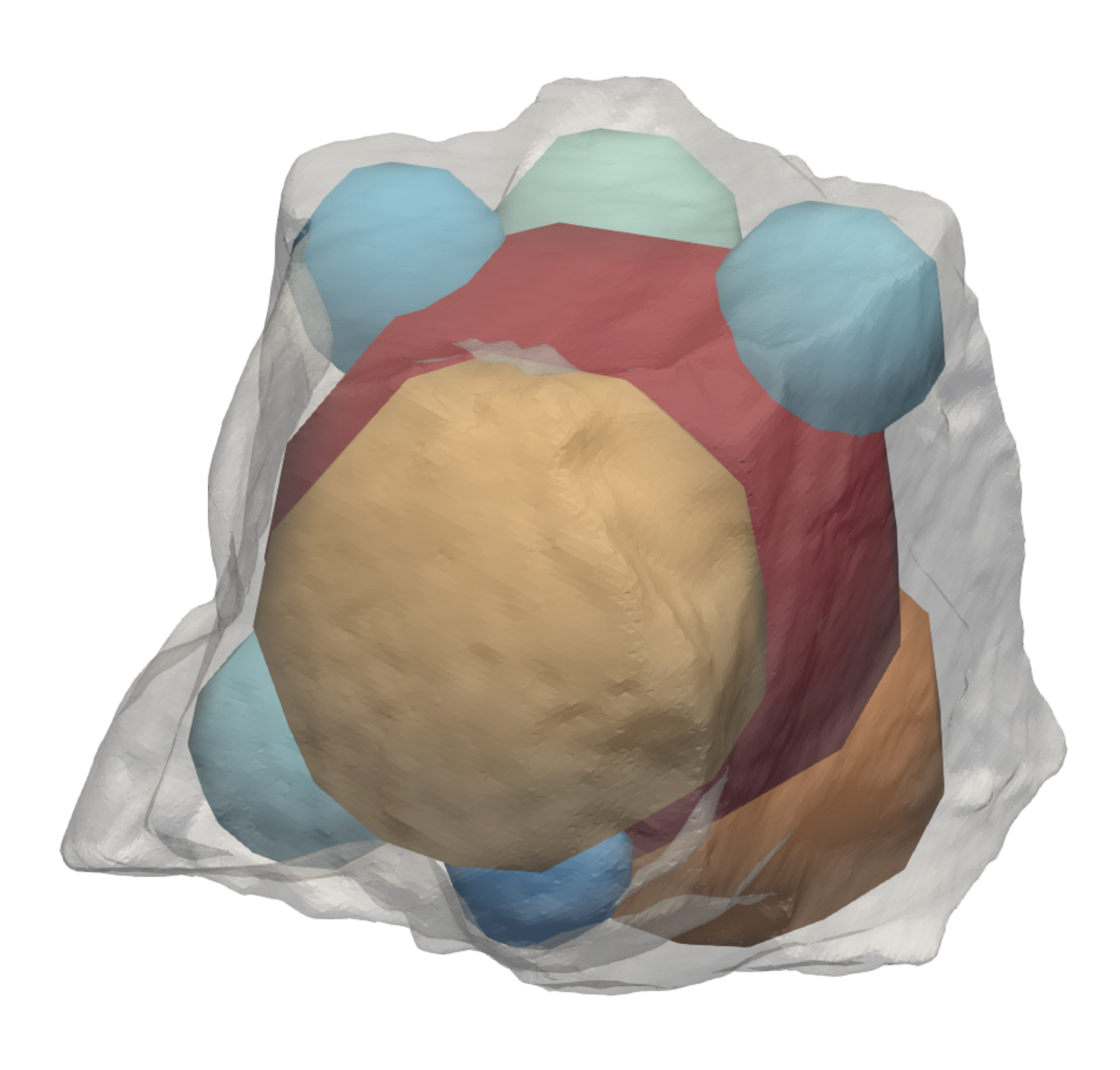}

    \end{minipage}
    \caption{A complex-shaped particle modeled with  $n=8$ and k=2 (left) and k=10 (right). Colors indicate the radii of the individual spheres. The particle corresponds to a realistic \textit{Hamburg sand} grain taken from the \textit{Sand Atlas} \cite{Milatz.2021} }
    \label{fig:keffect}
\end{figure}

In general, the optimal value of $k$ depends on the requirements and constraints of the user. A lower $k$ results in a smoother model surface, but requires a large number of spheres. A larger $k$ allows the spheres to be spaced further apart, capturing the overall bounding profile and extremities of the shape with fewer spheres, at the cost of increased artificial surface roughness. This tradeoff is not always captured by the Dice coefficient.  \cref{fig:keffect} shows the same particle modeled with 8 spheres for $k=2$ and $k=10$, yielding Dice coefficients of 0.811 and 0.799, respectively. However, visual inspection reveals that for $k=10$, the spheres are more widely distributed, representing the macroscopic structure more effectively.

\texttt{GEMSS} provides two options for computing the volume and the moment of inertia tensor: (a) from the voxelized multi-sphere model and (b) from the voxelized surface mesh of the target shape. 
\cref{fig:boxPrecision} shows the corresponding principal moments of inertia for the particle from \cref{fig:firstExample}.
\begin{figure}[htbp]
        \centering
\includegraphics[width=1.0\linewidth]{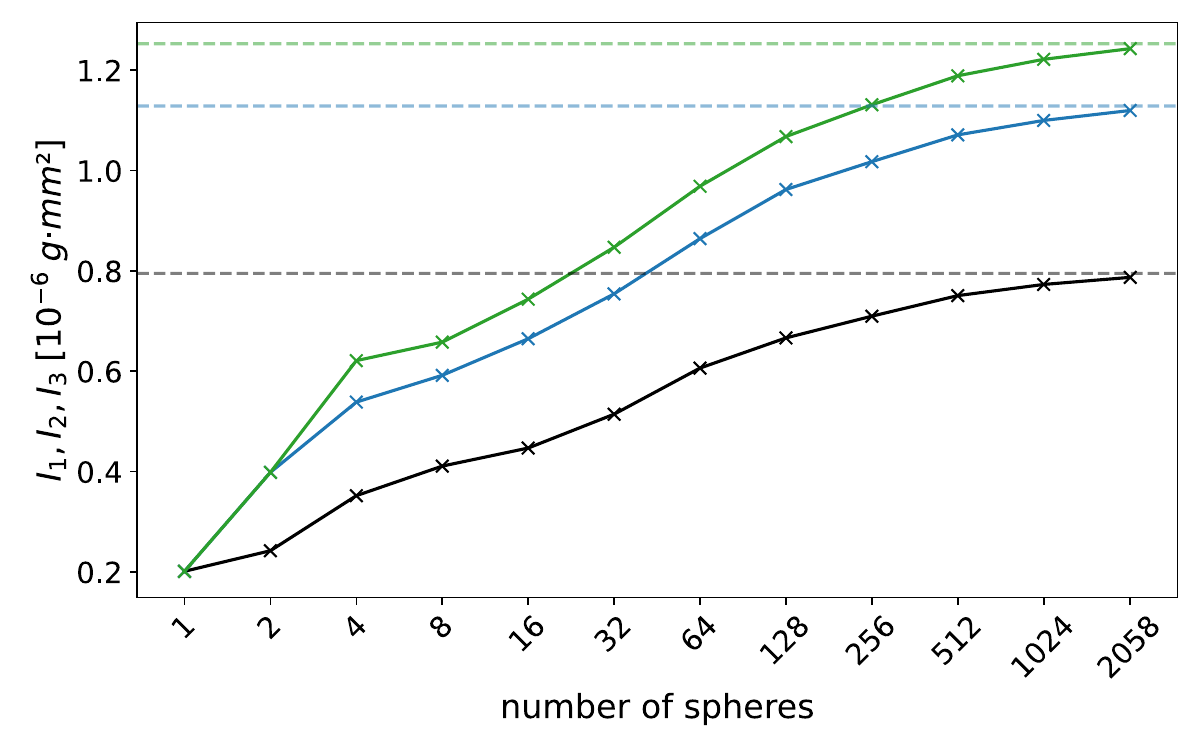}
\caption{Principal moments of inertia, $I_1\ge I_2\ge I_3$, for the particle shown in \cref{fig:firstExample}. The symbols show the data obtained from the multi-sphere model as functions of the number of spheres. The functions converge to the corresponding exact values obtained from the mesh data (dashed lines). 
}
\label{fig:boxMoments}
\end{figure}

In option (a), the moment of inertia tensor and the volume are computed from the multi-sphere representation itself. Consequently, the inertial properties are fully consistent with the particle used in the DEM simulation. However, because the particle is represented by a finite number of spheres, the resulting moments of inertia deviate from the exact values of the target shape.

In option (b), \texttt{GEMSS} computes the moment of inertia tensor directly from the target surface mesh. In this case, the moments of inertia match the target shape exactly, but are not fully consistent with the multi-sphere particle used in the DEM simulation.

\added{The choice between these two options depends on the specific requirements of the DEM application.  Option (b) is advantageous when computational constraints mandate a very low number of spheres, as Option (a) would yield severely inaccurate inertial properties in such cases.} As the number of spheres increases, the results obtained with option (a) converge to those of option (b).


\section{Impact}

\texttt{GEMSS} enables the use of dynamically changing particles in DEM simulations of granular materials within the simulation loop. This facilitates the simulation of key engineering processes such as comminution, grinding, and wear. \added{The host DEM framework handles the underlying physical models and provides the updated target shape, while GEMSS generates its multi-sphere model.} It has been integrated into MercuryDPM \cite{Weinhart.2020}, where multi-sphere particles are generated natively within the active simulation loop using a single command:\begin{lstlisting}
p1_= ClumpParticle::fromSTLFile(stlFilename, config); 
\end{lstlisting}
\added{A practical demonstration of this runtime integration and dynamic particle shape updating is available in the open-source MercuryDPM repository.}

\begin{figure}[htbp]
\centering
\includegraphics[width=1.0\linewidth]{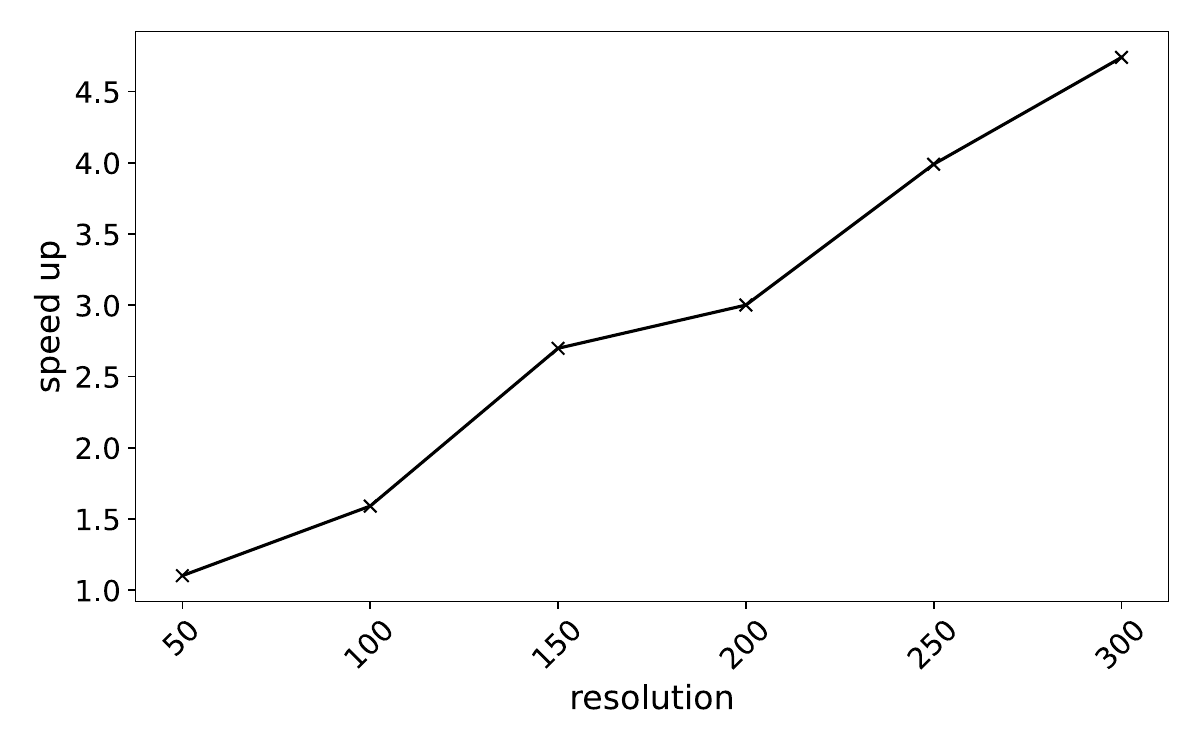}
\caption{Mean speed up of multi-sphere generation on 150 sand particles, in comparison to \cite{Multisphere.2026}. Both algorithms were benchmarked on the same machine to yield a relative comparison (executed on AMD Ryzen 5 5600G, Ubuntu 22.04.5 LTS, compiled with GCC 11 with -O3, using 12 threads).}

\label{fig:comparison}
\end{figure}

\texttt{GEMSS} reduces computational costs. Benchmarking 150 particles from the Sand Atlas \cite{Vego.2025, Milatz.2021, Vego.2023, Wiebicke.2017, Saadatfar.2012, Luijmes.2024} shows that \texttt{GEMSS} requires less execution time than the Python MSS implementation \cite{Multisphere.2026}, particularly at higher grid resolution. \cref{fig:comparison} shows the speed-up of \texttt{GEMSS} against Ref. \cite{Multisphere.2026} as a function of the resolution, specified by the user-defined parameter \texttt{div}, defined as the number of voxels along the shortest axis of the particle's axis-aligned bounding box.

\section{Conclusions}
\texttt{GEMSS} is an open-source, header-only C++ library for generating multi-sphere approximations of granular particles. By supporting both voxelized data and surface meshes, the software allows DEM and MD frameworks to load shapes directly during the simulation loop, offering a more integrated alternative to offline Python or MATLAB preprocessing scripts.

The library pairs the Multi-Sphere Shape (MSS) algorithm with the calculation of the volume, center of mass, and principal moments of inertia required for rigid-body dynamics, providing users with an efficient tool to  simulate complex, time-evolving particle shapes.

\bibliographystyle{elsarticle-num}
\bibliography{GEMSS_references}

\end{document}